\documentclass[prl,aps,twocolumn]{revtex4-1}
\usepackage{graphicx}
\usepackage{tabularx}

\begin{document}

\title{Gate defined zero- and one-dimensional confinement in bilayer graphene}

\author{Augustinus (Stijn) M. Goossens$^1$}
\email{a.m.goossens@tudelft.nl}
\author{Stefanie C.M. Driessen$^1$}
\author{Tim A. Baart$^1$}
\author{Kenji Watanabe$^2$}
\author{Takashi Taniguchi$^2$} 
\author{Lieven M.K. Vandersypen$^1$}

\affiliation{$^1$Kavli Institute of Nanoscience, Delft University of Technology, P.O. Box 5046, 2600 GA Delft, The Netherlands, $^2$Advanced Materials Laboratory, National Institute for Materials Science, 1-1 Namiki, Tsukuba, 305-0044, Japan}
\date{\today}

\begin{abstract}
We report on the fabrication and measurement of nanoscale devices based on bilayer graphene sandwiched between hexagonal boron nitride bottom and top gate dielectrics. The top gates are patterned such that constrictions and islands can be electrostatically induced by applying appropriate voltages to the gates. The high quality of the devices becomes apparent from conductance quantization in the constrictions at low temperature. The islands exhibit clear Coulomb blockade and single-electron transport.
\end{abstract}

\maketitle

\begin{figure}[b]
\begin{center}
\includegraphics[width=8.5cm]{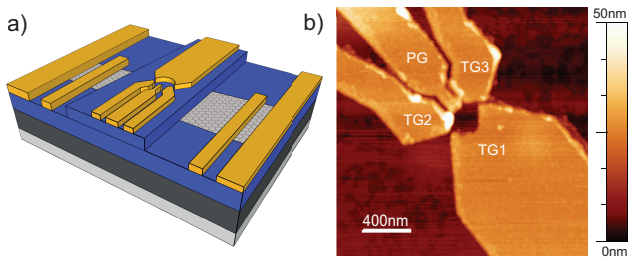}
\end{center}
\vspace*{-.5cm}
\caption{(a) Schematic layout of the sample. The lightgrey area indicates the heavily p-doped silicon wafer, dark grey the SiO$_{2}$ backgate dielectric, blue the hBN dielectrics and yellow Cr/Au electrodes and gates. In the actual sample, several top gate structures are present in between the contacts. (b) Atomic force microscope (AFM) height image of device A, with labeled top gates. Note that the plunger gate (PG) was not connected for device A.}
\vspace*{-.3cm}
\label{fig:overview}
\end{figure}

Confinement of charge carriers in graphene has been heavily investigated since graphene flakes on a substrate were first measured.\cite{geim_graphene:_2009, neto_electronic_2009} Nanopatterning graphene into nanoribbons or small islands has been a widely used strategy for both zero- and one-dimensional  confinement.\cite{ponomarenko_chaotic_2008, han_electron_2010, liu_electrostatic_2009, stampfer_tunable_2008, molitor_transport_2009} Unfortunately, as a result of the etching process that is used for nanopatterning, the edges of the devices are very rough. This edge disorder, aggravated by substrate-induced disorder, leads to rather irregular device behaviour. For instance no quantized conductance was observed in etched constrictions, the transparency of barriers varied non-monotonously with gate voltage, no atom-like shell-filling has been observed in quantum dots and in many cases quantum dots fell apart into multiple islands. 

Last year, suspended single layer graphene sheets narrowed by current induced heating showed quantized steps in conductance of $2e^2/h$ as the Fermi wavelength was varied.\cite{tombros_quantized_2011} Presumably these constrictions were not only narrow but also short, as observed in recent transmission electron microscope measurements \cite{barreiro_graphene_2012}, reducing the effects of edge disorder. However, the formation of these constrictions is hard to control, making it difficult to reproduce these results. Clean ribbon edges can be obtained reproducibly by unzipping carbon nanotubes, and a well-behaved quantum dot formed in such a nanoribbon was recently reported.\cite{wang_graphene_2011} Barriers were formed at metal Schottky contacts, but such barriers are not tunable, limiting follow-up work. Moreover, as the ribbons are not obtained lithographically but are dispersed from solution, they face many of the limitations of the carbon nanotube they originate from.

The ideal device would confine charge carriers in the bulk, far from (disordered) edges, have well controlled tunnel barriers, and enjoy all the design freedom offered by lithography. All these requirements can be satisfied using patterned electrostatic gates, provided a band gap is present. This calls for the use of bilayer graphene rather than monolayer graphene, as in bilayers a band gap can be induced by an electric field perpendicular to the layers.\cite{ mccann_low_2007,oostinga_gate-induced_2008,castro_biased_2007} An additional requirement for clean confinement is to minimize substrate-induced disorder. Substrate disorder can be eliminated in suspended devices with suspended top gates, as shown in recent work by
Allen et al.\cite{allen_gate_2012} However, it would be highly desirable to realize devices of comparable quality on a substrate, as this would facilitate integration of complex devices. Currently the cleanest gate dielectric available for graphene devices is hexagonal boron nitride (hBN), and mobilities reported on such substrates approach those of suspended devices.\cite{dean_boron_2010}

Here we report 1-dimensional and 0-dimensional electrostatic confinement in bilayer graphene sandwiched in hBN dielectrics. The devices have split top gates and a global back gate, which we bias so that a gap is opened. The Fermi level is tuned inside the gap in the regions below the top gates so that they become insulating. Depending on the combination of top gates that is biased, we form 1-dimensional channels or 0-dimensional islands.

A schematic and AFM image of a device is shown in Fig.~\ref{fig:overview}. We first deposit a $14$ nm thick hBN flake by mechanical exfoliation on a silicon wafer coated with a silicon oxide (SiO$_{2}$) layer of thickness $t_{SiO_{2}}=285$ nm. On top of the hBN we transfer a bilayer graphene flake ($\sim22\mu$m long and $\sim3\mu$m wide, its bilayer nature was confirmed by Raman spectroscopy) using a dry transfer method following the protocol of~\cite{dean_boron_2010} (at a temperature of $100~^{\circ}$C to remove any water absorbed on the surface of the graphene and hBN flakes). The sample was subsequently annealed in an oven at $400$ and $450~^{\circ}$C (Ar $2400$ sccm, H$_{2}$ $700$ sccm) to remove residues induced by the transfer process. Cr($5$ nm)/Au($95$ nm) electrodes are fabricated using electron-beam lithography (EBL). We annealed the samples again (same flow rate as the first annealing step, T = $300$, $350$ and $440~^{\circ}$C) to remove fabrication residues. This did not give the desired sample quality. Hence we applied the recently developed mechanical cleaning technique~\cite{goossens_mechanical_2012} followed by dry transfer of a $50$ nm thick hBN flake, which will act as top gate dielectric. In a two-step EBL process we deposited Cr/Au top gates. We defined several gate patterns between the contacts on this flake and report here on two 'quantum dot' top gate structures, one with a lithographic diameter of $320$ nm (device A) and one of $250$ nm (device B). The separation between the top gates that together define a barrier is less than $30$ nm. For device B, TG1 and TG2 were unintentionally connected.

This specific graphene sample was cooled down multiple times. The maximum overall mobility measured at T=$35$ mK was $~\sim36000$ cm$^2$/Vs (four-terminal configuration). The overall mobility measured during the last cool-down was much lower ($~\sim6000$ cm$^2$/Vs, two-terminal configuration). This degradation of electronic quality was caused by exposure to an electron beam during imaging in a scanning electron microscope. Nevertheless we believe that the graphene sandwiched in hBN remained of high quality. 

We set the back gate to a large negative voltage and tune the top gate to a voltage that compensates for the doping induced by the back gate. For typical values $V_{BG}=-50$ V and $V_{TG}=9$ V the displacement field is $D\sim0.6$ V/nm, which translates into a theoretically predicted band gap of $\sim50$ meV.\cite{mccann_low_2007,min_ab_2007, zhang_direct_2009} As in earlier work, the transport gap is substantially smaller~\cite{oostinga_gate-induced_2008, taychatanapat_electronic_2010} but still large enough to realize quantum confinement, as we will see.

\begin{figure}[b]
\begin{center}
\includegraphics[width=8.5cm]{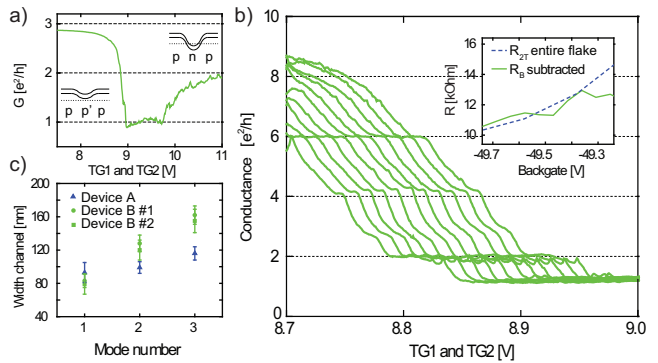}
\end{center}
\vspace*{-.5cm}
\caption{(a) Conductance versus the voltage on TG1 and TG2 (device B, $T=35$ mK, $V_{bias}=0.5$ mV, $V_{bg}=-49.74$ V). The other top gates of the device are set to $5$ V, far from pinch-off, so that we are dealing with only one constriction. All measurements presented in this paper are conducted in two-terminal DC voltage bias configuration. The conductance has been corrected for filter resistance and current amplifier input resistance. In the inset we show cartoons of the bandstructure for pp'p and pnp measurement regimes. Solid lines indicate valence and conduction band edges and dashed line the position of the Fermi level. (b) Top gate traces for device B at back gate voltages from $-49.26$ V (leftmost) to $-49.74$ V (rightmost) in $10$ equal steps ($V_{bias}=0.5$ mV, $T=35$ mK). The solid green line in the inset is the subtracted $R_{B}$. The dashed blue line is the two-terminal resistance of the graphene flake with all top gates at ground. (c) Channel width $W_n$ for modes $n=1$, 2 and 3 for device A (blue triangles) and for two measurements on device B (green circles and squares). The widths are calculated in the square-well potential approximation $W=\frac{n}{2}\lambda_{f}$. The Fermi wavelength at the $n$-th plateau is estimated via $\lambda_{f}=\frac{2\pi}{\sqrt{\pi C_{TG} \Delta V_{TG,n}/e}}$. $C_{TG}$ is the top gate capacitance per unit area, extracted from the slope of the position of the $\nu=4$ plateau in top gate voltage against magnetic field (Fig.~\ref{fig:bfield}b). $\Delta V_{TG,n}$ is the difference between the top gate voltage in the middle of the $n$-th plateau and the top gate voltage at charge neutrality.}
\vspace*{-.3cm}
\label{fig:qpc}
\end{figure}

In Fig.~\ref{fig:qpc}a we show a top gate trace at a large negative back gate voltage, taken at low temperature. Clearly visible is a region of suppressed conductance, with a remarkably clean transition region on the hole side, and a somewhat less clean transition and lower conductance on the electron side. This behavior can be expected since the leads are p-doped by the backgate, and a pnp-junction is formed on the electron side. The pnp-junction decreases the transparency of the device as the charge carriers have to Zener tunnel through the induced band gap. On the hole side the conductance suppression with top gate voltage is very smooth and well behaved. This is in strong contrast to similar top gate sweeps for graphene nanoribbons, which exhibit very irregular pinch-off characteristics when the Fermi energy is swept into the transport gap.\cite{han_electron_2010,molitor_transport_2009,liu_electrostatic_2009}

Zooming in on the steep flank of the pinch-off curve, we observe several plateaus in conductance, with a value that is independent of $V_{BG}$ (Fig.~\ref{fig:qpc}b). The traces have been corrected for filter resistance, current amplifier input resistance and for a background resistance ($R_{B}$) consisting of contact resistance, Maxwell spreading resistance and graphene lead resistance. The last two terms are dependent on back gate voltage. For each value of $V_{BG}$ we subtracted a value of $R_{B}$ comparable to the two-terminal resistance ($R_{2T}$) of the graphene flake with all top gates at ground (see inset fig.~\ref{fig:qpc}b), fine tuning $R_{B}$ such that the conductance $G$ at the upper plateau is $6e^2/h$.\footnote{Other choices of $R_B$ compatible with $R_{2T}$ gave either unphysical results, such as negative resistances, or quantized conductance values that we consider highly unlikely, such as plateaus at 8, 5 and 3 times $e^2/h$. Subtracting a parallel conductance of $e^2/h$, the value of the minimum conductance in the measurement, also gave unlikely conductance plateau values.} The other two plateaus consequently appear at $G=4e^2/h$ and $G=2e^2/h$. The same sequence of steps was observed for device A. (in device B, a less well developed feature can be seen just below $G=3e^2/h$, which does not appear in device A; its origin is unclear) This behavior is consistent with transport through one-dimensional ballistic channels and the formation of a quantum point contact.\cite{van_wees_quantized_1988, wharam_one-dimensional_1988} 

The $2e^2/h$ steps in conductance are surprising given that there is both spin and valley degeneracy in bulk bilayer graphene, so $4e^2/h$ steps are expected. The same observation was recently made on suspended gate-defined bilayer constrictions~\cite{allen_gate_2012} but it remains to be understood. Increasing the temperature to 440 mK did not change the general behavior. Lowering the bias to $50~\mu$V (lock-in measurement with an AC excitation of 10 $\mu$V) did not change the appearance of the plateaus either. Both a larger bias and a higher temperature smoothened out universal conductance fluctuations, as can be expected. 

We can estimate the width of the constriction from the position of the plateaus in top gate voltage with respect to the conductance minimum (Fig.~\ref{fig:qpc}c). We see that as subbands become occupied, the width of the constriction increases from $W_1\approx 90$ nm to $W_3\approx 120$ nm for device A and from $W_1 \approx 80$ to $W_3 \approx 160$ nm for device B, which is characteristic of a smooth confining potential (with $W_n$ the constriction width for the $n$-th subband). The lithographically defined separation between the respective top gates was less than $30$ nm for both devices. This implies that the channel extends below the top gates, which can be expected given the modest band gap induced underneath.

\begin{figure}[t]
\begin{center}
\includegraphics[width=8.5cm]{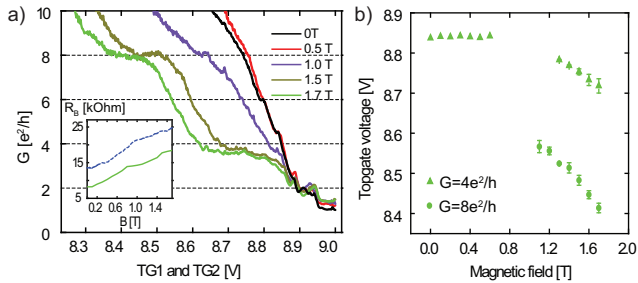}
\end{center}
\vspace*{-.5cm}
\caption{(a) Conductance  versus TG1 and TG2 voltage for device B at different magnetic fields (see legend, $T=440$ mK, $V_{bias}=150\mu$ V, $V_{bg}=-49.74$ V, $V_{PG}$ and $V_{TG3}=5$ V). The inset shows as a function of magnetic field the background resistances that we subtracted (solid green line) and the two-terminal resistance with all top gates at ground (dashed blue line, measured at $V_{bias}=500~\mu$V). In the low field regime the background resistance was determined by aligning the plateaus near G=$4e^2/h$ and in the high field regime by aligning the plateaus near G=$8e^2/h$ (b) Extracted top gate voltages at the plateau centres versus magnetic field.  By fitting a linear curve to the large field data we obtain $C_{TG1\&2}=\frac{4e^2}{h}\frac{dB}{dV_{TG,\nu=4}}\sim 1.04$ F/m$^2$. The value for $C_{TG1\&2}$ extracted from the $\nu=8$ data is very similar.} 
\vspace*{-.3cm}
\label{fig:bfield}
\end{figure}

In Fig.~\ref{fig:bfield}a we explore the influence of a perpendicular magnetic field on the 1D channels (device B). In the low-field regime, the plateau at $6e^2/h$ quickly disappears, but the plateaus at $4e^2/h$ and $2e^2/h$ remain visible. At large fields, plateaus in conductance develop at $4e^2/h$ ($\nu=4$) and $8e^2/h$ ($\nu=8$), typical values for the quantum Hall effect in bilayer graphene.\cite{novoselov_unconventional_2006} This transition from size confinement to magnetic confinement occurs when the cyclotron radius ($r_{c}$) is equal to or larger than $W_{n}/2$.\cite{van_wees_magnetoelectric_1988} By extrapolating the positions of the plateaus for $n=2$ and $\nu=4$, we can determine a cross-over magnetic field of $0.9\pm0.2$ T and estimate the size of the constriction based on $r_{c}$ (Fig.~\ref{fig:bfield}b). This gives $W_2=76\pm18$ nm. The agreement with the estimate based on the plateau positions ($\sim120$ nm) is better than a factor two.

\begin{figure}[t]
\begin{center}
\includegraphics[width=7.5cm]{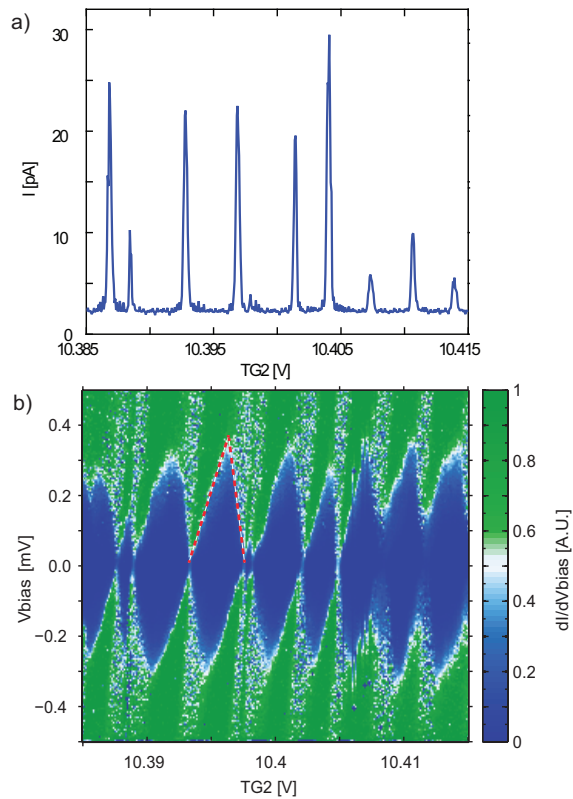}
\end{center}
\vspace*{-.5cm}
\caption{(a) Current versus the voltage on TG2 (device A, $V_{TG1}=V_{TG3}=10.1546$ V, $V_{BG}=-49.5$ V and $V_{bias}=50~\mu$V). (b) Differential conductance plotted in color scale versus gate voltage and source-drain bias (device A, $V_{TG1}=V_{TG3}=10.1546$ V, $V_{BG}=-49.5$ V). The red dashed lines indicate the outline of a diamond as used for the analysis of the capacitances. The tiny diamond in the middle is not included in the analysis.}
\vspace*{-.3cm}
\label{fig:cb1}
\end{figure}

\begin{figure}[t]
\begin{center}
\includegraphics[width=8.5cm]{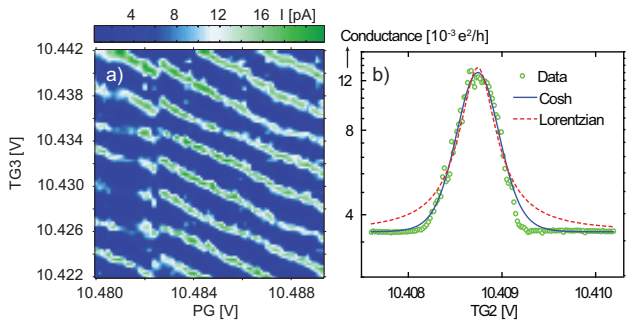}
\end{center}
\vspace*{-.5cm}
\caption{(a) Current plotted in color scale versus the voltage applied to PG and TG3 (device B, $V_{TG1\&2}=10.2548$ V, $V_{BG}=-49.38$ V and $V_{bias}=50~\mu$V ). Occasionally shifts occur such as the one at $V_{PG} \sim 10.4835$ V, which we attribute to a rearrangement of background charges in the substrate. (b) Zoom-in on a conductance peak plotted on a logarithmic scale along with fits by a Lorentzian and hyperbolic cosine function (device B, $V_{TG1\&2}=10.1907$ V, $V_{PG}=10.3880$ V, $V_{BG}=-49.86$ V and $V_{bias}=50~\mu$V). Fit formulas are
$G\propto \cosh^{-2}(\alpha \frac{V_{TG}-V_{TG,offset}}{2.5 kT})$ where $\alpha=\frac{2E_{C}}{\Delta V_{TG}}$ (the factor 2.5 assumes temperature exceeds level spacing) and $G\propto \frac{\Gamma}{(\alpha (V_{TG}-V_{TG,offset})/\hbar)^2+(\Gamma/2)^2}$, where $\Gamma$ is the coupling strength to the leads.}
\vspace*{-.3cm}
\label{fig:cb2}
\end{figure}

When we induce two barriers by appropriate gate voltages, the device behaviour changes drastically. Fig.~\ref{fig:cb1}a shows a gate voltage scan of TG2 while TG1 and TG3 are also biased (device A). We see sharp conductance peaks separated by regions of strongly suppressed conductance, which is characteristic of Coulomb blockade.\cite{kouwenhoven_few-electron_2001} As expected for Coulomb peaks, their position on one gate voltage axis varies smoothly (linearly) when another gate voltage is swept (Fig. ~\ref{fig:cb2}a). We note that the resistance in Coulomb blockade is orders of magnitude larger than the sum of the two barrier resistances, which saturates around $h/e^2$ (Fig. \ref{fig:qpc}), indicating each gate couples to both barriers.  Coulomb blockade is confirmed further by the diamond-shaped regions of suppressed conductance seen in a color plot of conductance versus gate voltage and bias voltage (Fig.~\ref{fig:cb1}b). 

The Coulomb peak spacing exhibits a standard deviation of $1.4$ mV, compared to an average peak spacing of $4$ mV (device B). Irregular peak spacings can, in principle, originate from quantized level spacing contributions.\cite{kouwenhoven_few-electron_2001} However, no excited state features are visible in the Coulomb diamonds so it is not clear that level spacing is responsible for the peak spacing variation. Instead the reason may be the breakdown of the constant-interaction model, for instance due to disorder-induced variations in the island size or position as charge carriers are added one by one, or a second island capacitively coupled to the main island.

From the Coulomb diamonds and peak spacing we can obtain information on the dimensions and location of the island. For the dataset of Fig.~\ref{fig:cb1}b (device A), the addition energy $E_{add}$ is $0.35\pm0.02$ meV. When we neglect orbital energies, $E_{add}=2E_{c}$, where $E_{c}$ is the charging energy. Using $E_{c}=\frac{e^2}{2C_{\Sigma}}$, we can calculate the total capacitance of the dot, $C_{\Sigma} = 0.46\pm0.03$ fF. From the slope of the Coulomb diamond edges, we find that $C_\Sigma$ is dominated by the source and drain capacitances ($C_{s}=0.10\pm0.024$ fF,  $C_{d}=0.30\pm0.033$ fF), which makes it difficult to estimate the island size from $E_{add}$ and $C_\Sigma$. 

Instead, we compare the measured top gate capacitances with top gate capacitances simulated using a 3D Poisson equation solver (Ansoft Maxwell). From the average Coulomb peak spacing $\Delta V_{TG2}$ we extract a capacitance $C_{TG2}\sim45$ aF. $C_{TG3}$, $C_{TG1}$ and $C_{PG}$ are comparable. The calculated capacitance between a circular island of $320$ nm in diameter and a metal plate with a $320$ nm hole, $50$ nm above the island, is $40$ aF, about as large as each measured top gate capacitance by itself. We thus infer that the island is formed in the central region uncovered by the top gates and extends underneath all the gates. This is consistent with the quantum point contacts extending underneath the split gates as discussed above (see also~\cite{allen_gate_2012}). Presumably lateral confinement is less tight in these devices than in GaAs split gate devices, due to the much smaller band gap.

Finally, we investigate whether the dot is weakly or strongly tunnel coupled to the leads by inspecting the line shape of the Coulomb peaks at low bias voltage. In Fig.~\ref{fig:cb2}b, we overlay on the data a fit by a hyperbolic cosine function and by a Lorentzian function. The former fits the data much better, indicating that the dot is in the weakly coupled regime, where temperature determines the linewidth rather than tunnel coupling to the reservoirs.\cite{beenakker_theory_1991} Doing the same analysis for 50 peaks we find that they are consistently in the weakly coupled regime. We extract an electron temperature of $69\pm14$ mK.

Concluding, we show evidence for one-dimensional and zero-dimensional confinement in substrate supported bilayer graphene split gate devices. Transport through a single-barrier device shows signs of conductance quantization and in double-barrier devices clear Coulomb blockade is observed. Further development of this platform will enable us to study the rich physics of quantum confined charge carriers in bilayer graphene. Questions to be addressed include how electron-electron and electron-phonon interactions, and spin and valley lifetimes are modified by the confinement.

We acknowledge useful discussions with A. Akhmerov, A. Barreiro, V. Calado, C. Dean, P. Kim, L. Kouwenhoven, N. Tombros, B. van Wees, M. Wimmer, A. Young and P. Zomer and financial support from the Foundation for Fundamental Research on Matter (FOM) and the European Research Council (ERC Starting Grant).

\bibliography{graphene}

\begin{thebibliography}{26}%
\makeatletter
\providecommand \@ifxundefined [1]{%
 \@ifx{#1\undefined}
}%
\providecommand \@ifnum [1]{%
 \ifnum #1\expandafter \@firstoftwo
 \else \expandafter \@secondoftwo
 \fi
}%
\providecommand \@ifx [1]{%
 \ifx #1\expandafter \@firstoftwo
 \else \expandafter \@secondoftwo
 \fi
}%
\providecommand \natexlab [1]{#1}%
\providecommand \enquote  [1]{``#1''}%
\providecommand \bibnamefont  [1]{#1}%
\providecommand \bibfnamefont [1]{#1}%
\providecommand \citenamefont [1]{#1}%
\providecommand \href@noop [0]{\@secondoftwo}%
\providecommand \href [0]{\begingroup \@sanitize@url \@href}%
\providecommand \@href[1]{\@@startlink{#1}\@@href}%
\providecommand \@@href[1]{\endgroup#1\@@endlink}%
\providecommand \@sanitize@url [0]{\catcode `\\12\catcode `\$12\catcode
  `\&12\catcode `\#12\catcode `\^12\catcode `\_12\catcode `\%12\relax}%
\providecommand \@@startlink[1]{}%
\providecommand \@@endlink[0]{}%
\providecommand \url  [0]{\begingroup\@sanitize@url \@url }%
\providecommand \@url [1]{\endgroup\@href {#1}{\urlprefix }}%
\providecommand \urlprefix  [0]{URL }%
\providecommand \Eprint [0]{\href }%
\providecommand \doibase [0]{http://dx.doi.org/}%
\providecommand \selectlanguage [0]{\@gobble}%
\providecommand \bibinfo  [0]{\@secondoftwo}%
\providecommand \bibfield  [0]{\@secondoftwo}%
\providecommand \translation [1]{[#1]}%
\providecommand \BibitemOpen [0]{}%
\providecommand \bibitemStop [0]{}%
\providecommand \bibitemNoStop [0]{.\EOS\space}%
\providecommand \EOS [0]{\spacefactor3000\relax}%
\providecommand \BibitemShut  [1]{\csname bibitem#1\endcsname}%
\let\auto@bib@innerbib\@empty
\bibitem [{\citenamefont {Geim}(2009)}]{geim_graphene:_2009}%
  \BibitemOpen
  \bibfield  {author} {\bibinfo {author} {\bibfnamefont {A.~K.}\ \bibnamefont
  {Geim}},\ }\href {\doibase 10.1126/science.1158877} {\bibfield  {journal}
  {\bibinfo  {journal} {Science}\ }\textbf {\bibinfo {volume} {324}},\ \bibinfo
  {pages} {1530 } (\bibinfo {year} {2009})}\BibitemShut {NoStop}%
\bibitem [{\citenamefont {Neto}\ \emph {et~al.}(2009)\citenamefont {Neto},
  \citenamefont {Guinea}, \citenamefont {Peres}, \citenamefont {Novoselov},\
  and\ \citenamefont {Geim}}]{neto_electronic_2009}%
  \BibitemOpen
  \bibfield  {author} {\bibinfo {author} {\bibfnamefont {A.~H.~C.}\
  \bibnamefont {Neto}}, \bibinfo {author} {\bibfnamefont {F.}~\bibnamefont
  {Guinea}}, \bibinfo {author} {\bibfnamefont {N.~M.~R.}\ \bibnamefont
  {Peres}}, \bibinfo {author} {\bibfnamefont {K.~S.}\ \bibnamefont
  {Novoselov}}, \ and\ \bibinfo {author} {\bibfnamefont {A.~K.}\ \bibnamefont
  {Geim}},\ }\href {\doibase 10.1103/RevModPhys.81.109} {\bibfield  {journal}
  {\bibinfo  {journal} {Rev. Mod. Phys.}\ }\textbf {\bibinfo {volume} {81}},\
  \bibinfo {pages} {109} (\bibinfo {year} {2009})}\BibitemShut {NoStop}%
\bibitem [{\citenamefont {Ponomarenko}\ \emph {et~al.}(2008)\citenamefont
  {Ponomarenko}, \citenamefont {Schedin}, \citenamefont {Katsnelson},
  \citenamefont {Yang}, \citenamefont {Hill}, \citenamefont {Novoselov},\ and\
  \citenamefont {Geim}}]{ponomarenko_chaotic_2008}%
  \BibitemOpen
  \bibfield  {author} {\bibinfo {author} {\bibfnamefont {L.~A.}\ \bibnamefont
  {Ponomarenko}}, \bibinfo {author} {\bibfnamefont {F.}~\bibnamefont
  {Schedin}}, \bibinfo {author} {\bibfnamefont {M.~I.}\ \bibnamefont
  {Katsnelson}}, \bibinfo {author} {\bibfnamefont {R.}~\bibnamefont {Yang}},
  \bibinfo {author} {\bibfnamefont {E.~W.}\ \bibnamefont {Hill}}, \bibinfo
  {author} {\bibfnamefont {K.~S.}\ \bibnamefont {Novoselov}}, \ and\ \bibinfo
  {author} {\bibfnamefont {A.~K.}\ \bibnamefont {Geim}},\ }\href {\doibase
  10.1126/science.1154663} {\bibfield  {journal} {\bibinfo  {journal}
  {Science}\ }\textbf {\bibinfo {volume} {320}},\ \bibinfo {pages} {356}
  (\bibinfo {year} {2008})}\BibitemShut {NoStop}%
\bibitem [{\citenamefont {Han}\ \emph {et~al.}(2010)\citenamefont {Han},
  \citenamefont {Brant},\ and\ \citenamefont {Kim}}]{han_electron_2010}%
  \BibitemOpen
  \bibfield  {author} {\bibinfo {author} {\bibfnamefont {M.~Y.}\ \bibnamefont
  {Han}}, \bibinfo {author} {\bibfnamefont {J.~C.}\ \bibnamefont {Brant}}, \
  and\ \bibinfo {author} {\bibfnamefont {P.}~\bibnamefont {Kim}},\ }\href
  {\doibase 10.1103/PhysRevLett.104.056801} {\bibfield  {journal} {\bibinfo
  {journal} {Phys. Rev. Lett.}\ }\textbf {\bibinfo {volume} {104}},\ \bibinfo
  {pages} {056801} (\bibinfo {year} {2010})}\BibitemShut {NoStop}%
\bibitem [{\citenamefont {Liu}\ \emph {et~al.}(2009)\citenamefont {Liu},
  \citenamefont {Oostinga}, \citenamefont {Morpurgo},\ and\ \citenamefont
  {Vandersypen}}]{liu_electrostatic_2009}%
  \BibitemOpen
  \bibfield  {author} {\bibinfo {author} {\bibfnamefont {X.}~\bibnamefont
  {Liu}}, \bibinfo {author} {\bibfnamefont {J.~B.}\ \bibnamefont {Oostinga}},
  \bibinfo {author} {\bibfnamefont {A.~F.}\ \bibnamefont {Morpurgo}}, \ and\
  \bibinfo {author} {\bibfnamefont {L.~M.~K.}\ \bibnamefont {Vandersypen}},\
  }\href {\doibase 10.1103/PhysRevB.80.121407} {\bibfield  {journal} {\bibinfo
  {journal} {Phys. Rev. B}\ }\textbf {\bibinfo {volume} {80}},\ \bibinfo
  {pages} {121407} (\bibinfo {year} {2009})}\BibitemShut {NoStop}%
\bibitem [{\citenamefont {Stampfer}\ \emph {et~al.}(2008)\citenamefont
  {Stampfer}, \citenamefont {Schurtenberger}, \citenamefont {Molitor},
  \citenamefont {G\"ttinger}, \citenamefont {Ihn},\ and\ \citenamefont
  {Ensslin}}]{stampfer_tunable_2008}%
  \BibitemOpen
  \bibfield  {author} {\bibinfo {author} {\bibfnamefont {C.}~\bibnamefont
  {Stampfer}}, \bibinfo {author} {\bibfnamefont {E.}~\bibnamefont
  {Schurtenberger}}, \bibinfo {author} {\bibfnamefont {F.}~\bibnamefont
  {Molitor}}, \bibinfo {author} {\bibfnamefont {J.}~\bibnamefont {G\"ttinger}},
  \bibinfo {author} {\bibfnamefont {T.}~\bibnamefont {Ihn}}, \ and\ \bibinfo
  {author} {\bibfnamefont {K.}~\bibnamefont {Ensslin}},\ }\href {\doibase
  10.1021/nl801225h} {\bibfield  {journal} {\bibinfo  {journal} {Nano Lett.}\
  }\textbf {\bibinfo {volume} {8}},\ \bibinfo {pages} {2378} (\bibinfo {year}
  {2008})}\BibitemShut {NoStop}%
\bibitem [{\citenamefont {Molitor}\ \emph {et~al.}(2009)\citenamefont
  {Molitor}, \citenamefont {Jacobsen}, \citenamefont {Stampfer}, \citenamefont
  {G\"uttinger}, \citenamefont {Ihn},\ and\ \citenamefont
  {Ensslin}}]{molitor_transport_2009}%
  \BibitemOpen
  \bibfield  {author} {\bibinfo {author} {\bibfnamefont {F.}~\bibnamefont
  {Molitor}}, \bibinfo {author} {\bibfnamefont {A.}~\bibnamefont {Jacobsen}},
  \bibinfo {author} {\bibfnamefont {C.}~\bibnamefont {Stampfer}}, \bibinfo
  {author} {\bibfnamefont {J.}~\bibnamefont {G\"uttinger}}, \bibinfo {author}
  {\bibfnamefont {T.}~\bibnamefont {Ihn}}, \ and\ \bibinfo {author}
  {\bibfnamefont {K.}~\bibnamefont {Ensslin}},\ }\href {\doibase
  10.1103/PhysRevB.79.075426} {\bibfield  {journal} {\bibinfo  {journal} {Phys.
  Rev. B}\ }\textbf {\bibinfo {volume} {79}},\ \bibinfo {pages} {075426}
  (\bibinfo {year} {2009})}\BibitemShut {NoStop}%
\bibitem [{\citenamefont {Tombros}\ \emph {et~al.}(2011)\citenamefont
  {Tombros}, \citenamefont {Veligura}, \citenamefont {Junesch}, \citenamefont
  {Guimaraes}, \citenamefont {{Vera-Marun}}, \citenamefont {Jonkman},\ and\
  \citenamefont {Van~Wees}}]{tombros_quantized_2011}%
  \BibitemOpen
  \bibfield  {author} {\bibinfo {author} {\bibfnamefont {N.}~\bibnamefont
  {Tombros}}, \bibinfo {author} {\bibfnamefont {A.}~\bibnamefont {Veligura}},
  \bibinfo {author} {\bibfnamefont {J.}~\bibnamefont {Junesch}}, \bibinfo
  {author} {\bibfnamefont {M.~H.~D.}\ \bibnamefont {Guimaraes}}, \bibinfo
  {author} {\bibfnamefont {I.~J.}\ \bibnamefont {{Vera-Marun}}}, \bibinfo
  {author} {\bibfnamefont {H.~T.}\ \bibnamefont {Jonkman}}, \ and\ \bibinfo
  {author} {\bibfnamefont {B.~J.}\ \bibnamefont {Van~Wees}},\ }\href {\doibase
  10.1038/nphys2009} {\bibfield  {journal} {\bibinfo  {journal} {Nat. Phys.}\
  }\textbf {\bibinfo {volume} {7}},\ \bibinfo {pages} {697} (\bibinfo {year}
  {2011})}\BibitemShut {NoStop}%
\bibitem [{\citenamefont {Barreiro}\ \emph {et~al.}(2012)\citenamefont
  {Barreiro}, \citenamefont {B\"orrnert}, \citenamefont {R\"ummeli},
  \citenamefont {B\"uchner},\ and\ \citenamefont
  {Vandersypen}}]{barreiro_graphene_2012}%
  \BibitemOpen
  \bibfield  {author} {\bibinfo {author} {\bibfnamefont {A.}~\bibnamefont
  {Barreiro}}, \bibinfo {author} {\bibfnamefont {F.}~\bibnamefont
  {B\"orrnert}}, \bibinfo {author} {\bibfnamefont {M.~H.}\ \bibnamefont
  {R\"ummeli}}, \bibinfo {author} {\bibfnamefont {B.}~\bibnamefont
  {B\"uchner}}, \ and\ \bibinfo {author} {\bibfnamefont {L.~M.~K.}\
  \bibnamefont {Vandersypen}},\ }\href {\doibase 10.1021/nl204236u} {\bibfield
  {journal} {\bibinfo  {journal} {Nano Lett.}\ }\textbf {\bibinfo {volume}
  {12}},\ \bibinfo {pages} {1873} (\bibinfo {year} {2012})}\BibitemShut
  {NoStop}%
\bibitem [{\citenamefont {Wang}\ \emph {et~al.}(2011)\citenamefont {Wang},
  \citenamefont {Ouyang}, \citenamefont {Jiao}, \citenamefont {Wang},
  \citenamefont {Xie}, \citenamefont {Wu}, \citenamefont {Guo},\ and\
  \citenamefont {Dai}}]{wang_graphene_2011}%
  \BibitemOpen
  \bibfield  {author} {\bibinfo {author} {\bibfnamefont {X.}~\bibnamefont
  {Wang}}, \bibinfo {author} {\bibfnamefont {Y.}~\bibnamefont {Ouyang}},
  \bibinfo {author} {\bibfnamefont {L.}~\bibnamefont {Jiao}}, \bibinfo {author}
  {\bibfnamefont {H.}~\bibnamefont {Wang}}, \bibinfo {author} {\bibfnamefont
  {L.}~\bibnamefont {Xie}}, \bibinfo {author} {\bibfnamefont {J.}~\bibnamefont
  {Wu}}, \bibinfo {author} {\bibfnamefont {J.}~\bibnamefont {Guo}}, \ and\
  \bibinfo {author} {\bibfnamefont {H.}~\bibnamefont {Dai}},\ }\href {\doibase
  10.1038/nnano.2011.138} {\bibfield  {journal} {\bibinfo  {journal} {Nat.
  Nano}\ }\textbf {\bibinfo {volume} {6}},\ \bibinfo {pages} {563} (\bibinfo
  {year} {2011})}\BibitemShut {NoStop}%
\bibitem [{\citenamefont {{McCann}}\ \emph {et~al.}(2007)\citenamefont
  {{McCann}}, \citenamefont {Abergel},\ and\ \citenamefont
  {Fal'ko}}]{mccann_low_2007}%
  \BibitemOpen
  \bibfield  {author} {\bibinfo {author} {\bibfnamefont {E.}~\bibnamefont
  {{McCann}}}, \bibinfo {author} {\bibfnamefont {D.~S.}\ \bibnamefont
  {Abergel}}, \ and\ \bibinfo {author} {\bibfnamefont {V.~I.}\ \bibnamefont
  {Fal'ko}},\ }\href {\doibase 10.1140/epjst/e2007-00229-1} {\bibfield
  {journal} {\bibinfo  {journal} {The European Physical Journal - Special
  Topics}\ }\textbf {\bibinfo {volume} {148}},\ \bibinfo {pages} {91} (\bibinfo
  {year} {2007})}\BibitemShut {NoStop}%
\bibitem [{\citenamefont {Oostinga}\ \emph {et~al.}(2008)\citenamefont
  {Oostinga}, \citenamefont {Heersche}, \citenamefont {Liu}, \citenamefont
  {Morpurgo},\ and\ \citenamefont {Vandersypen}}]{oostinga_gate-induced_2008}%
  \BibitemOpen
  \bibfield  {author} {\bibinfo {author} {\bibfnamefont {J.~B.}\ \bibnamefont
  {Oostinga}}, \bibinfo {author} {\bibfnamefont {H.~B.}\ \bibnamefont
  {Heersche}}, \bibinfo {author} {\bibfnamefont {X.}~\bibnamefont {Liu}},
  \bibinfo {author} {\bibfnamefont {A.~F.}\ \bibnamefont {Morpurgo}}, \ and\
  \bibinfo {author} {\bibfnamefont {L.~M.~K.}\ \bibnamefont {Vandersypen}},\
  }\href {\doibase 10.1038/nmat2082} {\bibfield  {journal} {\bibinfo  {journal}
  {Nat. Mater.}\ }\textbf {\bibinfo {volume} {7}},\ \bibinfo {pages} {151}
  (\bibinfo {year} {2008})}\BibitemShut {NoStop}%
\bibitem [{\citenamefont {Castro}\ \emph {et~al.}(2007)\citenamefont {Castro},
  \citenamefont {Novoselov}, \citenamefont {Morozov}, \citenamefont {Peres},
  \citenamefont {dos Santos}, \citenamefont {Nilsson}, \citenamefont {Guinea},
  \citenamefont {Geim},\ and\ \citenamefont {Neto}}]{castro_biased_2007}%
  \BibitemOpen
  \bibfield  {author} {\bibinfo {author} {\bibfnamefont {E.~V.}\ \bibnamefont
  {Castro}}, \bibinfo {author} {\bibfnamefont {K.~S.}\ \bibnamefont
  {Novoselov}}, \bibinfo {author} {\bibfnamefont {S.~V.}\ \bibnamefont
  {Morozov}}, \bibinfo {author} {\bibfnamefont {N.~M.~R.}\ \bibnamefont
  {Peres}}, \bibinfo {author} {\bibfnamefont {J.~M. B.~L.}\ \bibnamefont {dos
  Santos}}, \bibinfo {author} {\bibfnamefont {J.}~\bibnamefont {Nilsson}},
  \bibinfo {author} {\bibfnamefont {F.}~\bibnamefont {Guinea}}, \bibinfo
  {author} {\bibfnamefont {A.~K.}\ \bibnamefont {Geim}}, \ and\ \bibinfo
  {author} {\bibfnamefont {A.~H.~C.}\ \bibnamefont {Neto}},\ }\href {\doibase
  10.1103/PhysRevLett.99.216802} {\bibfield  {journal} {\bibinfo  {journal}
  {Phys. Rev. Lett.}\ }\textbf {\bibinfo {volume} {99}},\ \bibinfo {pages}
  {216802} (\bibinfo {year} {2007})}\BibitemShut {NoStop}%
\bibitem [{\citenamefont {Allen}\ \emph {et~al.}(2012)\citenamefont {Allen},
  \citenamefont {Martin},\ and\ \citenamefont {Yacoby}}]{allen_gate_2012}%
  \BibitemOpen
  \bibfield  {author} {\bibinfo {author} {\bibfnamefont {M.~T.}\ \bibnamefont
  {Allen}}, \bibinfo {author} {\bibfnamefont {J.}~\bibnamefont {Martin}}, \
  and\ \bibinfo {author} {\bibfnamefont {A.}~\bibnamefont {Yacoby}},\ }\href
  {http://arxiv.org/abs/1202.0820} {\bibfield  {journal} {\bibinfo  {journal}
  {{arXiv:1202.0820}}\ } (\bibinfo {year} {2012})}\BibitemShut {NoStop}%
\bibitem [{\citenamefont {Dean}\ \emph {et~al.}(2010)\citenamefont {Dean},
  \citenamefont {Young}, \citenamefont {Meric}, \citenamefont {Lee},
  \citenamefont {Wang}, \citenamefont {Sorgenfrei}, \citenamefont {Watanabe},
  \citenamefont {Taniguchi}, \citenamefont {Kim}, \citenamefont {Shepard},\
  and\ \citenamefont {Hone}}]{dean_boron_2010}%
  \BibitemOpen
  \bibfield  {author} {\bibinfo {author} {\bibfnamefont {C.}~\bibnamefont
  {Dean}}, \bibinfo {author} {\bibfnamefont {A.}~\bibnamefont {Young}},
  \bibinfo {author} {\bibfnamefont {I.}~\bibnamefont {Meric}}, \bibinfo
  {author} {\bibfnamefont {C.}~\bibnamefont {Lee}}, \bibinfo {author}
  {\bibfnamefont {L.}~\bibnamefont {Wang}}, \bibinfo {author} {\bibfnamefont
  {S.}~\bibnamefont {Sorgenfrei}}, \bibinfo {author} {\bibfnamefont
  {K.}~\bibnamefont {Watanabe}}, \bibinfo {author} {\bibfnamefont
  {T.}~\bibnamefont {Taniguchi}}, \bibinfo {author} {\bibfnamefont
  {P.}~\bibnamefont {Kim}}, \bibinfo {author} {\bibfnamefont {K.}~\bibnamefont
  {Shepard}}, \ and\ \bibinfo {author} {\bibfnamefont {J.}~\bibnamefont
  {Hone}},\ }\href {\doibase 10.1038/nnano.2010.172} {\bibfield  {journal}
  {\bibinfo  {journal} {Nat. Nano}\ }\textbf {\bibinfo {volume} {5}},\ \bibinfo
  {pages} {722} (\bibinfo {year} {2010})}\BibitemShut {NoStop}%
\bibitem [{\citenamefont {Goossens}\ \emph {et~al.}(2012)\citenamefont
  {Goossens}, \citenamefont {Calado}, \citenamefont {Barreiro}, \citenamefont
  {Watanabe}, \citenamefont {Taniguchi},\ and\ \citenamefont
  {Vandersypen}}]{goossens_mechanical_2012}%
  \BibitemOpen
  \bibfield  {author} {\bibinfo {author} {\bibfnamefont {A.~M.}\ \bibnamefont
  {Goossens}}, \bibinfo {author} {\bibfnamefont {V.~E.}\ \bibnamefont
  {Calado}}, \bibinfo {author} {\bibfnamefont {A.}~\bibnamefont {Barreiro}},
  \bibinfo {author} {\bibfnamefont {K.}~\bibnamefont {Watanabe}}, \bibinfo
  {author} {\bibfnamefont {T.}~\bibnamefont {Taniguchi}}, \ and\ \bibinfo
  {author} {\bibfnamefont {L.~M.~K.}\ \bibnamefont {Vandersypen}},\ }\href
  {\doibase doi:10.1063/1.3685504} {\bibfield  {journal} {\bibinfo  {journal}
  {Appl. Phys. Lett.}\ }\textbf {\bibinfo {volume} {100}},\ \bibinfo {pages}
  {073110} (\bibinfo {year} {2012})}\BibitemShut {NoStop}%
\bibitem [{\citenamefont {Min}\ \emph {et~al.}(2007)\citenamefont {Min},
  \citenamefont {Sahu}, \citenamefont {Banerjee},\ and\ \citenamefont
  {{MacDonald}}}]{min_ab_2007}%
  \BibitemOpen
  \bibfield  {author} {\bibinfo {author} {\bibfnamefont {H.}~\bibnamefont
  {Min}}, \bibinfo {author} {\bibfnamefont {B.}~\bibnamefont {Sahu}}, \bibinfo
  {author} {\bibfnamefont {S.~K.}\ \bibnamefont {Banerjee}}, \ and\ \bibinfo
  {author} {\bibfnamefont {A.~H.}\ \bibnamefont {{MacDonald}}},\ }\href
  {\doibase 10.1103/PhysRevB.75.155115} {\bibfield  {journal} {\bibinfo
  {journal} {Phys. Rev. B}\ }\textbf {\bibinfo {volume} {75}},\ \bibinfo
  {pages} {155115} (\bibinfo {year} {2007})}\BibitemShut {NoStop}%
\bibitem [{\citenamefont {Zhang}\ \emph {et~al.}(2009)\citenamefont {Zhang},
  \citenamefont {Tang}, \citenamefont {Girit}, \citenamefont {Hao},
  \citenamefont {Martin}, \citenamefont {Zettl}, \citenamefont {Crommie},
  \citenamefont {Shen},\ and\ \citenamefont {Wang}}]{zhang_direct_2009}%
  \BibitemOpen
  \bibfield  {author} {\bibinfo {author} {\bibfnamefont {Y.}~\bibnamefont
  {Zhang}}, \bibinfo {author} {\bibfnamefont {T.}~\bibnamefont {Tang}},
  \bibinfo {author} {\bibfnamefont {C.}~\bibnamefont {Girit}}, \bibinfo
  {author} {\bibfnamefont {Z.}~\bibnamefont {Hao}}, \bibinfo {author}
  {\bibfnamefont {M.~C.}\ \bibnamefont {Martin}}, \bibinfo {author}
  {\bibfnamefont {A.}~\bibnamefont {Zettl}}, \bibinfo {author} {\bibfnamefont
  {M.~F.}\ \bibnamefont {Crommie}}, \bibinfo {author} {\bibfnamefont {Y.~R.}\
  \bibnamefont {Shen}}, \ and\ \bibinfo {author} {\bibfnamefont
  {F.}~\bibnamefont {Wang}},\ }\href {\doibase 10.1038/nature08105} {\bibfield
  {journal} {\bibinfo  {journal} {Nature}\ }\textbf {\bibinfo {volume} {459}},\
  \bibinfo {pages} {820} (\bibinfo {year} {2009})}\BibitemShut {NoStop}%
\bibitem [{\citenamefont {Taychatanapat}\ and\ \citenamefont
  {{Jarillo-Herrero}}(2010)}]{taychatanapat_electronic_2010}%
  \BibitemOpen
  \bibfield  {author} {\bibinfo {author} {\bibfnamefont {T.}~\bibnamefont
  {Taychatanapat}}\ and\ \bibinfo {author} {\bibfnamefont {P.}~\bibnamefont
  {{Jarillo-Herrero}}},\ }\href {\doibase 10.1103/PhysRevLett.105.166601}
  {\bibfield  {journal} {\bibinfo  {journal} {Phys. Rev. Lett.}\ }\textbf
  {\bibinfo {volume} {105}},\ \bibinfo {pages} {166601} (\bibinfo {year}
  {2010})}\BibitemShut {NoStop}%
\bibitem [{Note1()}]{Note1}%
  \BibitemOpen
  \bibinfo {note} {Other choices of $R_B$ compatible with $R_{2T}$ gave either
  unphysical results, such as negative resistances, or quantized conductance
  values that we consider highly unlikely, such as plateaus at 8, 5 and 3 times
  $e^2/h$. Subtracting a parallel conductance of $e^2/h$, the value of the
  minimum conductance in the measurement, also gave unlikely conductance
  plateau values.}\BibitemShut {Stop}%
\bibitem [{\citenamefont {Van~Wees}\ \emph
  {et~al.}(1988{\natexlab{a}})\citenamefont {Van~Wees}, \citenamefont
  {Van~Houten}, \citenamefont {Beenakker}, \citenamefont {Williamson},
  \citenamefont {Kouwenhoven}, \citenamefont {van~der Marel},\ and\
  \citenamefont {Foxon}}]{van_wees_quantized_1988}%
  \BibitemOpen
  \bibfield  {author} {\bibinfo {author} {\bibfnamefont {B.~J.}\ \bibnamefont
  {Van~Wees}}, \bibinfo {author} {\bibfnamefont {H.}~\bibnamefont
  {Van~Houten}}, \bibinfo {author} {\bibfnamefont {C.~W.~J.}\ \bibnamefont
  {Beenakker}}, \bibinfo {author} {\bibfnamefont {J.~G.}\ \bibnamefont
  {Williamson}}, \bibinfo {author} {\bibfnamefont {L.~P.}\ \bibnamefont
  {Kouwenhoven}}, \bibinfo {author} {\bibfnamefont {D.}~\bibnamefont {van~der
  Marel}}, \ and\ \bibinfo {author} {\bibfnamefont {C.~T.}\ \bibnamefont
  {Foxon}},\ }\href {\doibase 10.1103/PhysRevLett.60.848} {\bibfield  {journal}
  {\bibinfo  {journal} {Phys. Rev. Lett.}\ }\textbf {\bibinfo {volume} {60}},\
  \bibinfo {pages} {848} (\bibinfo {year} {1988}{\natexlab{a}})}\BibitemShut
  {NoStop}%
\bibitem [{\citenamefont {Wharam}\ \emph {et~al.}(1988)\citenamefont {Wharam},
  \citenamefont {Thornton}, \citenamefont {Newbury}, \citenamefont {Pepper},
  \citenamefont {Ahmed}, \citenamefont {Frost}, \citenamefont {Hasko},
  \citenamefont {Peacock}, \citenamefont {Ritchie},\ and\ \citenamefont
  {Jones}}]{wharam_one-dimensional_1988}%
  \BibitemOpen
  \bibfield  {author} {\bibinfo {author} {\bibfnamefont {D.~A.}\ \bibnamefont
  {Wharam}}, \bibinfo {author} {\bibfnamefont {T.~J.}\ \bibnamefont
  {Thornton}}, \bibinfo {author} {\bibfnamefont {R.}~\bibnamefont {Newbury}},
  \bibinfo {author} {\bibfnamefont {M.}~\bibnamefont {Pepper}}, \bibinfo
  {author} {\bibfnamefont {H.}~\bibnamefont {Ahmed}}, \bibinfo {author}
  {\bibfnamefont {J.~E.~F.}\ \bibnamefont {Frost}}, \bibinfo {author}
  {\bibfnamefont {D.~G.}\ \bibnamefont {Hasko}}, \bibinfo {author}
  {\bibfnamefont {D.~C.}\ \bibnamefont {Peacock}}, \bibinfo {author}
  {\bibfnamefont {D.~A.}\ \bibnamefont {Ritchie}}, \ and\ \bibinfo {author}
  {\bibfnamefont {G.~A.~C.}\ \bibnamefont {Jones}},\ }\href {\doibase
  10.1088/0022-3719/21/8/002} {\bibfield  {journal} {\bibinfo  {journal} {J.
  Phys. C: Solid State Phys.}\ }\textbf {\bibinfo {volume} {21}},\ \bibinfo
  {pages} {L209} (\bibinfo {year} {1988})}\BibitemShut {NoStop}%
\bibitem [{\citenamefont {Novoselov}\ \emph {et~al.}(2006)\citenamefont
  {Novoselov}, \citenamefont {{McCann}}, \citenamefont {Morozov}, \citenamefont
  {Fal'ko}, \citenamefont {Katsnelson}, \citenamefont {Zeitler}, \citenamefont
  {Jiang}, \citenamefont {Schedin},\ and\ \citenamefont
  {Geim}}]{novoselov_unconventional_2006}%
  \BibitemOpen
  \bibfield  {author} {\bibinfo {author} {\bibfnamefont {K.~S.}\ \bibnamefont
  {Novoselov}}, \bibinfo {author} {\bibfnamefont {E.}~\bibnamefont {{McCann}}},
  \bibinfo {author} {\bibfnamefont {S.~V.}\ \bibnamefont {Morozov}}, \bibinfo
  {author} {\bibfnamefont {V.~I.}\ \bibnamefont {Fal'ko}}, \bibinfo {author}
  {\bibfnamefont {M.~I.}\ \bibnamefont {Katsnelson}}, \bibinfo {author}
  {\bibfnamefont {U.}~\bibnamefont {Zeitler}}, \bibinfo {author} {\bibfnamefont
  {D.}~\bibnamefont {Jiang}}, \bibinfo {author} {\bibfnamefont
  {F.}~\bibnamefont {Schedin}}, \ and\ \bibinfo {author} {\bibfnamefont
  {A.~K.}\ \bibnamefont {Geim}},\ }\href {\doibase 10.1038/nphys245} {\bibfield
   {journal} {\bibinfo  {journal} {Nat. Phys.}\ }\textbf {\bibinfo {volume}
  {2}},\ \bibinfo {pages} {177} (\bibinfo {year} {2006})}\BibitemShut {NoStop}%
\bibitem [{\citenamefont {Van~Wees}\ \emph
  {et~al.}(1988{\natexlab{b}})\citenamefont {Van~Wees}, \citenamefont
  {Kouwenhoven}, \citenamefont {van Houten}, \citenamefont {Beenakker},
  \citenamefont {Mooij}, \citenamefont {Foxon},\ and\ \citenamefont
  {Harris}}]{van_wees_magnetoelectric_1988}%
  \BibitemOpen
  \bibfield  {author} {\bibinfo {author} {\bibfnamefont {B.~J.}\ \bibnamefont
  {Van~Wees}}, \bibinfo {author} {\bibfnamefont {L.~P.}\ \bibnamefont
  {Kouwenhoven}}, \bibinfo {author} {\bibfnamefont {H.}~\bibnamefont {van
  Houten}}, \bibinfo {author} {\bibfnamefont {C.~W.~J.}\ \bibnamefont
  {Beenakker}}, \bibinfo {author} {\bibfnamefont {J.~E.}\ \bibnamefont
  {Mooij}}, \bibinfo {author} {\bibfnamefont {C.~T.}\ \bibnamefont {Foxon}}, \
  and\ \bibinfo {author} {\bibfnamefont {J.~J.}\ \bibnamefont {Harris}},\
  }\href {\doibase 10.1103/PhysRevB.38.3625} {\bibfield  {journal} {\bibinfo
  {journal} {Phys. Rev. B}\ }\textbf {\bibinfo {volume} {38}},\ \bibinfo
  {pages} {3625} (\bibinfo {year} {1988}{\natexlab{b}})}\BibitemShut {NoStop}%
\bibitem [{\citenamefont {Kouwenhoven}\ \emph {et~al.}(2001)\citenamefont
  {Kouwenhoven}, \citenamefont {Austing},\ and\ \citenamefont
  {Tarucha}}]{kouwenhoven_few-electron_2001}%
  \BibitemOpen
  \bibfield  {author} {\bibinfo {author} {\bibfnamefont {L.~P.}\ \bibnamefont
  {Kouwenhoven}}, \bibinfo {author} {\bibfnamefont {D.~G.}\ \bibnamefont
  {Austing}}, \ and\ \bibinfo {author} {\bibfnamefont {S.}~\bibnamefont
  {Tarucha}},\ }\href {\doibase 10.1088/0034-4885/64/6/201} {\bibfield
  {journal} {\bibinfo  {journal} {Rep. Prog. Phys.}\ }\textbf {\bibinfo
  {volume} {64}},\ \bibinfo {pages} {701} (\bibinfo {year} {2001})}\BibitemShut
  {NoStop}%
\bibitem [{\citenamefont {Beenakker}(1991)}]{beenakker_theory_1991}%
  \BibitemOpen
  \bibfield  {author} {\bibinfo {author} {\bibfnamefont {C.~W.~J.}\
  \bibnamefont {Beenakker}},\ }\href {\doibase 10.1103/PhysRevB.44.1646}
  {\bibfield  {journal} {\bibinfo  {journal} {Phys. Rev. B}\ }\textbf {\bibinfo
  {volume} {44}},\ \bibinfo {pages} {1646} (\bibinfo {year}
  {1991})}\BibitemShut {NoStop}%
\end{thebibliography}%


\end{document}